\documentclass[english]{article}
\usepackage{mathptmx}

\usepackage[T1]{fontenc}
\usepackage[utf8x]{inputenc}
\usepackage{array}
\usepackage{float}
\usepackage{multirow}
\usepackage{graphicx}
\usepackage[authoryear]{natbib}
\PassOptionsToPackage{normalem}{ulem}
\usepackage{ulem}

\makeatletter

\providecommand{\tabularnewline}{\\}
\newcommand{\lyxdot}{.}

\usepackage{booktabs}

\makeatother

\usepackage{babel}
\begin{document}

\title{Understanding Civil War Violence through Military Intelligence: Mining
Civilian Targeting Records from the Vietnam War}

\author{Rex W. Douglass%
\thanks{I thank David Madden, Josh Martin, Walter Fick, and Roxanna Ramzipoor
for research assistance, and the United States National Archives staff,
particularly Richard Boylan and Lynn Goodsell for generously sharing
their time and expertise. I am grateful for comments from Erik Gartzke,
Joanne Gowa, Kristen Harkness, Stathis Kalyvas, Chris Kennedy, Matthew
Kocher, Alex Lanoszka, John Lindsay, David Meyer, Kris Ramsay, Jacob
Shapiro, Tom Scherer, members of the Empirical Studies of Conflict
Group, the Yale Program on Order, Conflict, and Violence, the UCSD
Cross Domain Deterrence Group, and two anonymous reviewers. This research
was supported, in part, by the U.S. Department of Defenses Minerva
Research Initiative through the Air Force Office of Scientific Research,
grant \#FA9550-09-1-0314.%
}}

\date{June 2015%
\thanks{Forthcoming as a chapter in C.A. Anderton and J. Brauer, eds., Economic
Aspects of Genocides, Mass Atrocities, and Their Prevention. New York:
Oxford University Press.%
}}
\maketitle
\begin{abstract}
Military intelligence is underutilized in the study of civil war violence.
Declassified records are hard to acquire and difficult to explore
with the standard econometrics toolbox. I investigate a contemporary
government database of civilians targeted during the Vietnam War.
The data are detailed, with up to 45 attributes recorded for 73,712
individual civilian suspects. I employ an unsupervised machine learning
approach of cleaning, variable selection, dimensionality reduction,
and clustering. I find support for a simplifying typology of civilian
targeting that distinguishes different kinds of suspects and different
kinds targeting methods. The typology is robust, successfully clustering
both government actors and rebel departments into groups that mirror
their known functions. The exercise highlights methods for dealing
with high dimensional found conflict data. It also illustrates how
aggregating measures of political violence masks a complex underlying
empirical data generating process as well as a complex institutional
reporting process.
\end{abstract}

\section{Introduction}

Civil wars blur the line between civilians and combatants. This is
the fundamental problem for governments that must separate rebels
from innocents and for civilians wanting to remain neutral and safe.
Military and police forces expend enormous resources attempting to
identify and eliminate specific individuals fighting for violent groups.
What do those programs look like from the inside? How do they pick
their targets? How effective are they, and how are their costs distributed
across both civilian and rebel supporters?

Targeting programs are necessarily secretive, which makes government
records hard to come by. When available, they tend to consist of unwieldy,
unstructured, and often undigitized intelligence dossiers, which makes
quantitative analysis an expensive proposition.%
\footnote{Two prime examples include the East German Stasi files which came
into public stewardship, and the Guatemalan National Police Archive
which is now in the charge of government and international humanitarian
agencies as part of a truth and reconciliation effort \citep{aguirre_silence_2013}.
The records are comprehensive but unstructured and will require a
major effort to analyze once the all of the raw documents are digitized
\citep{price_statistical_2009}.%
} Because of these problems, nearly all existing studies of targeting
programs are either qualitative in nature \citep{moyar_phoenix_1997,comber_malayas_2008,natapoff_snitching:_2009}
or depend on data developed from nongovernment sources like interviews,
surveys, and news reports \citep{ball_bosnian_2007,silva_violent_2009}.%
\footnote{For some of the issues and methodology of working with retrospective
sources see \citep{price_big_2014,seybolt_counting_2013}.%
} When declassified government records are available, they tend to
be at the event level, like attacks \citep{biddle_testing_2012,berman_can_2011}
or air operations \citep{lyall_bombing_2014} without details on the
victims. The rare exception are peacetime police records like Stop
and Frisk data from New York, which provide information on both demographic
details about the suspect and details of the altercation \citep{gelman_analysis_2007}.

I analyze an electronic database of civilian targeting efforts created
during the Vietnam War. The data are extensive, covering 73,712 individual
rebel suspects from the perspective of the government's police and
military operations. The database contains detailed information on
both the victims and the operations targeting them.%
\footnote{With few exceptions, the database and most declassified intelligence
products from the Vietnam War have remained unused. This is partially
because working with found data is technically challenging, requiring
extensive cleaning and documentation— and partially because the tools
for such data are only now gaining popularity in the social sciences.%
} At issue are two central questions.

The first is descriptive: How exactly does a civilian targeting program
work in practice? All civilian targeting programs are secretive, but
the Phoenix Program is uniquely surrounded in historical controversy.
During and immediately after the war critics and proponents debated
whether it was a broad intelligence and policing effort or simply
a punitive assassination program \citep{colby_lost_1989}. Since the
war, the discussion has turned to whether the Phoenix Program achieved
its aims of neutralizing high ranking targets \citep{thayer_war_1985}.
More recently, it has been asked why the program caught some suspects
while letting others escape, and what implications that might have
for civilians deciding whether to join rebellion \citep{kalyvas_how_2007}.

The second is theoretical: How should we conceptualize civilian targeting?
Is there a simpler topology we can use to classify violence against
civilians in terms of the kinds of victims or the kinds of methods
employed? Much of the work on civilian targeting either explicitly
or implicitly dissagregates along the severity of targeting, treating
killings as distinct from arrests because they are theoretically different
or easier to document. Others divide targeting along how the victims
are selected, particularly whether they were individually singled
out or targeted as part of a larger group (Kalyvas 2006). Is there
a principled and data driven way to categorize and describe civilian
targeting?

Providing an answer to both questions requires a highly inductive
and multivariate approach. The overall analytical strategy is familiar
in the machine learning literature. I start with a database I did
not create and for which I have partial, incomplete documentation.
Using clues like patterns of missingness, I determine that there are
related groups of attributes and records drawn from completely different
sub-populations of suspects. I then show how to prioritize and reduce
the over forty available attributes for each suspect to a manageable
core group of key facts. I finally perform dimensionality reduction
on those facts, finding useful and easy to understand dimensions on
which both suspects and targeting tactics varied-- providing an answer
to both how we should conceptualize civilian targeting and how the
Phoenix Program operated.

In the course of conducting the analysis, I intend the chapter to
serve as a primer for handling large, multivariate, found conflict
data. An increasing number of scholars works with observational data
that they did not create, and over which they had no control. They
spend extraordinary effort on model specification for a suspected
empirical data generating process, but typically pay little attention
to the institutional data generating process underlying the reporting
and recording of facts. The Phoenix Program database provides an example
of how working with found data is at best an adversarial relationship.
This poses unique problems for inference, but I demonstrate some of
the growing number of tools that help to tackle these kinds of problems.

The road map of the chapter is as follows. The second section provides
an empirical background for civilian targeting during the Vietnam
War. The third section provides a broad overview of the data created
to track that targeting. A detailed examination of the database reveals
undocumented heterogeneity between different kinds of records, groups
of related attributes, and a ranking attributes' importance, pinpointing
exactly where to start the analysis.

Section four develops a taxonomy of victims. Dimensionality reduction
on a set of key attributes reveals two key differences between suspects:
(1) priority, those who the government wishes it could target and
those it targets in practice; and (2) severity, how violent the suspect's
final fate, ranging from voluntary defection, to arrest or capture,
and at the extreme, death.

Section five develops a taxonomy of tactics. Dimensionality reduction
on just observations where the government carried out an operation
(killings and arrests) reveals differences between tactics along:
(1) priority, but also premeditation, how specifically the individual
was targeted prior to the arrest or killing; and (2) domain, whether
a suspect was targeted in operations with some connection to policing
and intelligence resources, or whether the suspect was targeted by
third party forces and then retroactively reported.

A final section concludes with broader implications for security studies
and suggests avenues for future research.

\section{The Empirical Background: War in South Vietnam and the Phoenix Program}

In the later years of the Vietnam War, the South Vietnamese government
went on the offensive against the nonmilitary members and supporters
of the rebel opposition. While politically effective, the 1968 Tet-Offensive
was a military setback for irregular forces that shifted the main
military threat from the organized rural insurgency to the incoming
conventional forces from North Vietnam. The government took advantage
of this shift by pushing out into rural areas, projecting its power
with a wide range of police, militia, military, and special forces
units. With the expansion of government control and a flood of funds,
logistical support, and U.S. advisers, the government set out to map,
monitor, and police its civilian population at an industrial scale.
Goals were set out to dismantle the political opposition with widespread
arrests, killings, and induced defections.

The Phoenix Program (Phung Hoang) was created in 1968 to coordinate
this initiative — which was, in reality, fractured across dozens of
separate intelligence and police initiatives across South Vietnam.
Its substance as an institution included national guidelines that
influenced targeting goals and reporting, as well as physical offices
containing advisors who coordinated and distributed information on
suspects. While the program was not responsible for directly acting
against suspects, it was instrumental in bringing together and documenting
all of the scattered existing efforts \citep{moyar_phoenix_1997}.
It was in operation until it collapsed during the Eastertide invasion
by North Vietnam at the end of 1972.%
\footnote{It began out of a regional coordination effort called Intelligence
Coordination and Exploitation for the attack on the VCI (ICEX) in
July 1967. By 1968, Phung Hoang Committees were established 44 provinces
and 228 districts.%
}

The process by which the program collected information was scattered
and disjoint. Intelligence and Operations Coordinating Centers (IOCCs)
maintained Lists of Communist Offenders and sometimes detailed maps
of hamlets, with names of occupants and photographs.%
\footnote{Example members of a typical district IOCC team included Village Chiefs,
Deputy Village Chiefs for Security, Village Military Affairs Commissioners,
Village National Police Chiefs, Popular Forces Platoon Leaders, Hamlet
Chiefs, and more.%
} Military units like the 1st Infantry Brigade, 5th Infantry Division
(mechanized) sometimes formed special teams from its military intelligence
detachment to coordinate with and make up for weak IOCCs in their
area, maintaining their own card file in addition to the Local List
of Communist Offenders.%
\footnote{Military Assistance Advisory Group. Vietnam Lessons Learned No. 80:
US Combat Forces in Support of Pacification, 29 June 1970. Saigon,
Vietnam: Headquarters, U. S. Army Section, Military Assistance Advisory
Group, 1970-06-29 http://cgsc.contentdm.oclc.org/u?/p4013coll11,1524.%
} Additionally, village, district, and province chiefs often maintained
their own parallel intelligence nets and records.%
\footnote{``Phung Hoang Review'', December 1970. Pg 11 from VIETCONG INFRASTRUCTURE
NEUTRALIZATION SYSTEMS (VCINS), no date, Folder 065, US Marine Corps
History Division Vietnam War Documents Collection, The Vietnam Center
and Archive, Texas Tech University. Accessed 17 Apr. 2015. <http://www.vietnam.ttu.edu/virtualarchive/items.php?item=1201065056>.%
}

In January of 1969, the Office of the Secretary through the Advanced
Research Projects Agency (the predecessor to DARPA) began Project
VICEX to develop a country-wide information reporting system to coordinate
and allow comparative analysis of IOCC information.%
\footnote{``Org and Mission,'' April 1969, Phung Hoang Directorate, Records
of the Office of Civil Operations for Rural Development Support (CORDS),
General Records, 1967-1971; Record Group 472.3.10. National Archives
at College Park, College Park, MD. ARC Identifier: 4495500.%
} Biographical data on suspects and neutralizations were entered into
the national database at district and province IOCCs. Enumerators
were trained with coding guidelines for converting dossiers and neutralization
reports into a standardized format.%
\footnote{Reference Copy of Technical Documentation for Accessioned Electronic
Records, National Police Infrastructure Analysis Subsystem (NPIASS)
I \& II Master Files, Record Group 472 Records of the U.S. Forces
in Southeast Asia. Electronic Records Division, U.S. National Archives
and Records Administration, College Park, Md.%
} The data were then passed up to the Phung Hoang Directorate and from
there to the National Police Command Data Management Center, where
they were recorded using punch-cards before being entered onto magnetic
tape with an IBM360 mainframe.

\section{Overview of the Targeting Database}

The United States preserved an electronic copy of the final targeting
database called the National Police Infrastructure Analysis Subsystem
II (NPIASS-II).%
\footnote{United States Military Assistance Command Vietnam/Civil Operations
Rural Development Support (MACORDS) National Police Infrastructure
Analysis Subsystem II (NPIASS II), 1971-1973. File Number 3-349-79-992-D.
Created by the Military Assistance Command/Civil Operations and Rural
Development Support-Research and Analysis (MACORDS-RA). U.S. Military
Assistance Command/Civil Operations Rural Development Support.%
} It contains records from July 1970 to December 1972. This is the
later period of the war, following the Tet Offensive, the GVN and
the U.S. counter attack, and the period of U.S. withdrawal. It contains
data on 73,717 individuals (rows) that I refer to as suspects and
who serve as the unit of observation. It contains 45 attributes (columns)
that provide information on each suspect about their biography, job,
details of operations targeted against them, and their final disposition.%
\footnote{The total number of attributes is higher if multipart attributes are
disaggregated or if low to no variance attributes are included.%
} The attributes are of mixed types including numerical, nominal, dates,
and nested lookup codes such as locations (e.g. region->province->district->village)
and rebel jobs (e.g. National Liberation Front->Liberation Woman's
Association->Personnel). The structure of the database is outlined
in Table 1.

\begin{table}[H]
\hfill{}%
\begin{tabular}{|>{\centering}m{0.5in}|>{\centering}m{1in}|>{\centering}m{1in}|>{\centering}m{1.25in}|}
\cline{2-4} 
\multicolumn{1}{>{\centering}m{0.5in}|}{} & Record Type & Status & Outcome\tabularnewline
\hline 
\multirow{4}{0.5in}{Suspects (73,712)} & \multirow{2}{1in}{Neutralization Record (48,074)} & \multirow{3}{1in}{Neutralized (49,756)} & Killed (15,438)\tabularnewline
\cline{4-4} 
 &  &  & Captured (22,215)\tabularnewline
\cline{2-2} \cline{4-4} 
 & \multirow{2}{1in}{Biographical Record (25,638)} &  & Defector (12,103)\tabularnewline
\cline{3-4} 
 &  & At Large (23,943) & At Large (23,943)\tabularnewline
\hline 
\end{tabular}\hfill{}

\protect\caption{Structure of the National Police Infrastructure Analysis Subsystem
II (NPIASS-II) database. The unit of observation (rows) are individual
suspects. The potential attributes (columns) are available in blocks
depending on the kind of record and outcome.}
\end{table}

The first-order task with found data such as these is to verify the
structure of the database. Available codebooks often do not document
important details, and when they do, the documentation is often at
odds with how the database was used in practice. I take a two-pronged
approach. First, I point to archival evidence about the genesis of
and day-to-day use of the program. In particular, I located detailed
coding instructions for a precursor database, the VCI Neutralization
and Identification Information System (VCINIIS), which appears to
share most if not all of the same properties of the final NPIASS-II
database.%
\footnote{``VCI Neutralization and Identification Information System (VCINIIS)
Reporting and Coordination Procedures.'' Folder: \textquotedbl{}1603-03A
Operational Aids, 1969,\textquotedbl{} ARC Identifier: 5958372, Administrative
and Operational Records, compiled 05/1967 - 1970, documenting the
period 1966 - 1970, HMS Entry Number(s): A1 724, Record Group 472:
Records of the U.S. Forces in Southeast Asia, 1950 - 1976. National
Archives at College Park, College Park, MD.%
} Second, I apply a machine learning approach to the structure of the
database overall, not just the tabular values of individual variables.
Patterns of missing values, the meaning of different variables, and
heterogeneity between different kinds of observations are all targets
of inquiry.

One revelation from the documentation of the predecessor system is
that the database combines two different kinds of records: those entered
while a suspect was still at large (a biographical record) and those
entered after a suspect had already been killed, captured, or defected
(a neutralization record). The two kinds of records resulted from
different worksheets, with separate text examples, and coding guidelines.
Neutralization records appear to have recorded the Phoenix Program
as it actually happened on the ground. Two thirds of suspects, 48,074
(65\%), were entered in as neutralization records. Biographical records
were a kind of growing wish list where analysts bothered to digitize
information from the much larger pool of suspects with dossiers, on
blacklists, or reported in the Political Order of Battle. A third
of suspects, 25,638 (35\%), were entered this way.

The second important division in the database is along the fate of
each suspect, as still at large, killed, captured, or defector (rallier).
All bibliographic records start off as at large suspects. In rare
cases (6.6\%), some were updated to show the suspect had later been
neutralized, though there are sufficient irregularities to suspect
that some of these are actually coding errors.

The Phoenix Program was not, at least predominately, an assassination
program. The dominant form of neutralization was arrests/captures
(45\%). A smaller share (24\%) of neutralizations were suspects that
turned themselves in, defecting. Killings made up only a third of
neutralizations documented by the database (31\%).

\subsection{Example Narrative and Coding}

To illustrate the kinds of information available (and the kinds of
facts that are omitted), I provide the following comparison of a suspect's
record in the database with their detailed interrogation report. The
record was deanonymized by manually comparing data fields against
reported details in declassified interrogation reports from the Combined
Military Interrogation Center (CMIC).%
\footnote{``VCI of the Soc Trang Province Party Committee,'' March 1971, Folder
11, Box 09, Douglas Pike Collection: Unit 05 - National Liberation
Front, The Vietnam Center and Archive, Texas Tech University. Accessed
9 Apr. 2015. http://www.vietnam.ttu.edu/virtualarchive/items.php?item=2310911004.%
} Below is a brief narrative of a suspect's life, career, circumstance
of defection, and immediate aftermath drawn from his interrogation
report and with facts corroborated by the NPIASS-II database underlined
in the text.

In \uline{1929}, a \uline{man} named To Van Xiem was born in
Thai Binh Province (North Vietnam). He attended four years of village
school and worked on his parents' farm until January 1950 where he
joined the Viet Minh. He was a probationary member beginning June
1950 and became a \uline{full party member} in 1951. Over the next
sixteen years he was promoted or reassigned to multiple roles within
North Vietnam until February 25, 1966, when he and fifty-six other
civilian political cadre infiltrated South Vietnam, arriving in Tay
Ninh Province. In July of that year, he was assigned as a cadre of
the Ca Mau Province Party Committee Farmers' Association Section.
In March of 1967 he was reassigned to the same position in Soc Trang
Province. In September 1969 he became a member of the \uline{Current
Affairs Section of the Farmers' Association Section}, living in \uline{Ba
Xuyen Province}. He was a Buddhist, middle class farmer, married twice
with several children. Citing increasing hardships, limited rations,
and almost total government control of the province, he \uline{defected}
at \uline{Xuyen District, Ba Xuyen Province} on \uline{March
21, 1971}. He was previously \uline{unknown} to security forces,
not listed on any blacklist, and his identity was confirmed by \uline{confession}.
Four days later, \uline{March, 25, 1971}, his neutralization was
entered into the national database. He was assigned a VCI serial number
\uline{41-100825}, indicating a new, previously unlisted \uline{neutralization
record}. He was interrogated July 25, 1971 by the Combined Military
Interrogation Center (CMIC) in Saigon where he was assigned a CMIC
number of 0297-71. Details from his interrogation produced a 30 page
report that documented the names and details of 52 other individuals
as well as a number of regional organizations.

\subsection{The Attributes}

The full list of 45 attributes are provided below in Table 2. Attributes
are arranged into groups, determined using a combination of descriptions
from the codebooks and an unsupervised clustering method described
fully below. I have further aggregated the groups into four broad
concepts; attributes that are always available, typically only available
for biographical records, typically only available for neutralization
records, and only available for neutralization records of suspects
who defected or were captured.

\begin{table}[H]
\hfill{}%
\begin{tabular}{cccc}
\hline 
Almost Always Available & Biographical Record & Neutralization Record & Captured/Defector\tabularnewline
\hline 
Serial Number & At Large$^{\ddagger}$ & Killed/Capt./Defect. & Detention Facility\tabularnewline
Job & Birth Place & Action Force & \tabularnewline
 & Bio Process. Date & ID Source & Arrest Level\tabularnewline
Echelon & Bio Info Date & Neut. Process. Date & Arrest Serial\tabularnewline
Sex & Dossier Location & Neut. Action Date & Arrest Year\tabularnewline
Black List &  & Neut. Location & \tabularnewline
Party Membership & Photo &  & Sentence Process. Date\tabularnewline
Area of Operation & Prints & Specific Target & Sentence Date\tabularnewline
Priority A/B$^{*}$ & Arrest Order & Operation Level & Sentence Code\tabularnewline
Record Updates & Address & IOCC Involvement & Sentence Location\tabularnewline
 & Confirmation &  & \tabularnewline
Age &  &  & Release Process. Date\tabularnewline
 &  &  & Release Action Date\tabularnewline
 &  &  & Release Code\tabularnewline
 &  &  & Release Location\tabularnewline
 &  &  & \tabularnewline
 &  &  & Arrest Forwarding\tabularnewline
 &  &  & Forw. Process. Date\tabularnewline
 &  &  & Forw. Action Date\tabularnewline
 &  &  & Forw. Location\tabularnewline
\hline 
\multicolumn{4}{c}{{\small{}$^{*}$Imputed from official position Greenbook. $^{\ddagger}$Mutually
exclusive and so merged with Killed/Capt./Defector}}\tabularnewline
\end{tabular}\hfill{}

\protect\caption{Attributes for each suspect in the NPIASS-II database. }
\end{table}

\subsection{Grouping Attributes and Records}

The codebooks provide descriptions of each attribute but they omit
important details such as how missing values are handled. Every observation
has at least some missing attributes and nearly 50\% of cells are
empty. Some of the patterns are self-explanatory, e.g. there will
be no information about sentencing if the suspect is still at large.
In other cases, missingness is more subtle, e.g. information on the
suspect's age is sometimes missing for suspects who were killed in
the field without questioning. In all cases there appears to be a
combination of missing at random and undocumented structure. This
ambiguity and apparent latent structure suggests applying a machine
learning approach to learning how attributes are related to one another.

I frame this as a blockclustering problem where the task is to simultaneously
find $r$ groups of attributes and $k$ groups of observations that
are similar in terms of missingness. Let an $I\times Q$ binary matrix,
$X^{NA}$, represent the missingness for each individual $i$ and
attribute $q$, shown on the left in Figure 1. The task is to decompose
this matrix into a version sorted by row and column into homogenous
blocks $\hat{X}{}^{NA}$, shown on the right side of Figure 1, and
a smaller $r$ by $k$ binary matrix of row and column clusters.

I employ an unsupervised bi-clustering algorithm, the Bernoulli Latent
Block Model \citep{govaert_clustering_2003}.%
\footnote{ Implemented in the R package Blockcluster \citep{bhatia_blockcluster:_2014}.%
} The model is fit with a wide range of possible row and column cluster
counts, and the final model is selected with the best fit according
to the integrated complete likelihood (ICL) \citep{biernacki_assessing_2000}.

\begin{figure}[H]
\hfill{}\includegraphics[width=4.5in]{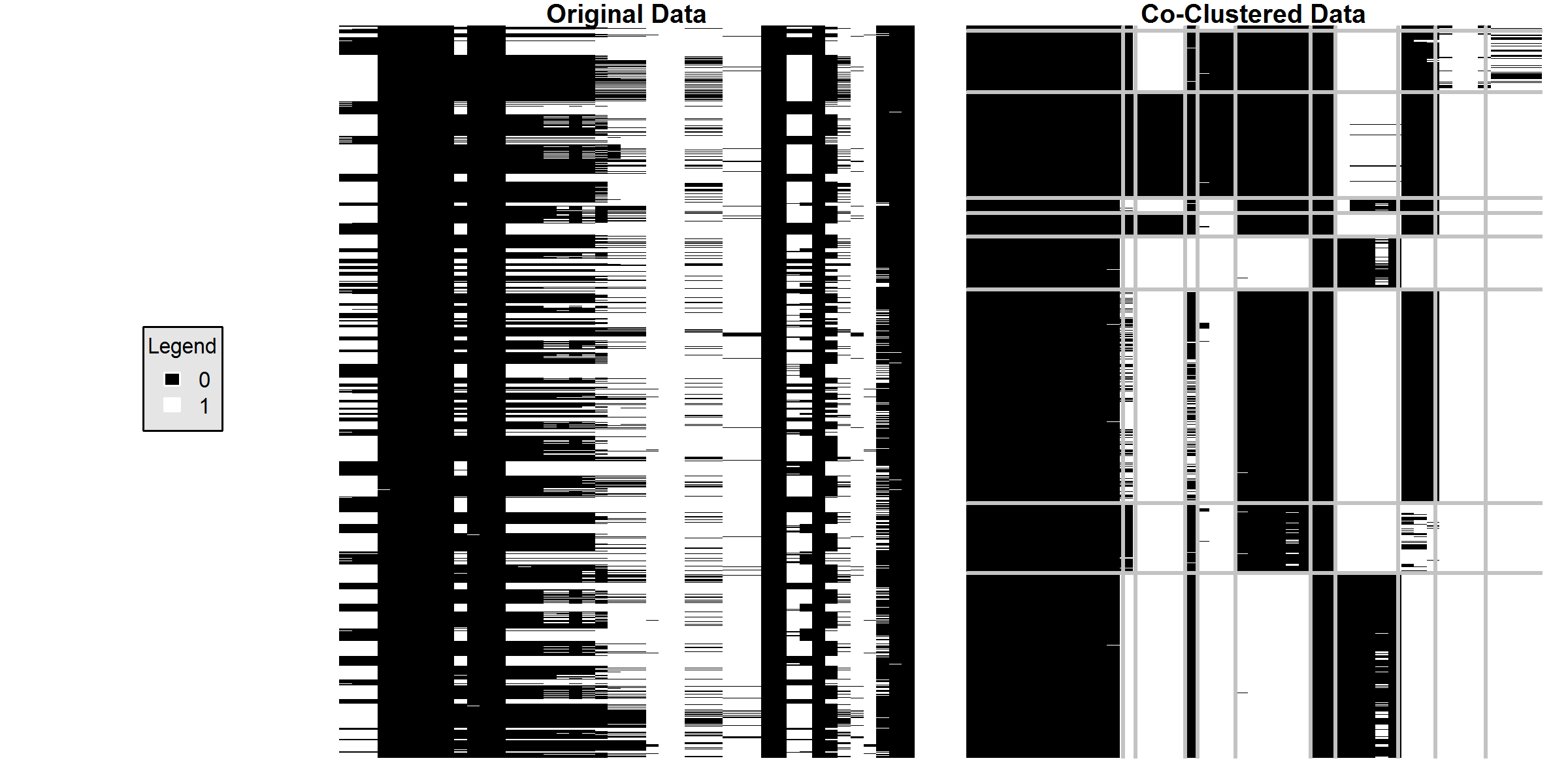}\hfill{}

\protect\caption{Block clustering of missing values in NPIASS-II into 11 groups of
attributes and 9 groups of observations. True values (white) indicate
a missing value.}
\end{figure}

The structure of missingness in the NPIASS-II database is best explained
by 11 clusters of attributes and 9 clusters of observations (ICL$=-260339.9$).
For substantive reasons discussed next, I further split off dossier
related attributes as a separate cluster bringing the total to 12.
These are the low level groups of attributes shown in Table 2.

In almost every case, the method has recovered known groups of variables
as detailed in the codebook. In a few cases it has correctly identified
an attribute as belonging to a different group despite a misleading
original variable name. The six record clusters recover the undocumented
split between biographical records and neutralization records, the
existence of neutralizations with additional follow up information
about sentencing and release, and the existence of a small number
of biographical records that were updated with neutralizations. Since
these record clusters might have further substantive implications,
I include record cluster as an additional attribute in the analysis
below.

Why devote special care to analysis of missingness? In this instance,
motivation comes from a discovered detail with major substantive implications
for the only other recent study to use this database. \citet{kalyvas_how_2007}
ask whether a suspect with weak evidence against them was more or
less likely to be neutralized than a suspect with strong evidence
against them. They focus on the dossier attribute called ``confirmed''
which is when a suspect is identified by three different sources or
one irrefutable source. They find that confirmed suspects, with presumably
more evidence against them, were actually much less likely to be neutralized
than unconfirmed suspects. If true, this would have a troubling implication
that the guilty can be much safer than the innocent from government
targeting in an insurgency.

In fact, this counter-intuitive result turns entirely on a technical
detail; ``confirmed'' and a few other dossier related attributes
were stored as a 0/1 bit flag in original IBM360 system. When they
were converted to modern formats, 0 values were converted to ``No''
instead of missing and so all 48,074 neutralization records were accidentally
counted as unconfirmed neutralizations. Confirmation may have played
a role in neutralization, but the database does not record it in a
way that makes the desired comparison possible.

This is a tricky mistake to catch, an undocumented difference in record
types plus an undocumented imputation of values. However, it pops
out in this analysis of missingness since there appears to be at least
two different kinds of records mixed together and all of the other
dossier attributes tend to be missing for neutralization records.
It also pops out in the analysis of attribute importance and interaction
that I turn to next.

\subsection{Variable Selection}

Each suspect in the database has potentially over forty known facts
about them. If you were forced to describe an observation to someone
else, which fact would you start with?%
\footnote{Put another way, suspects are situated in some high dimensional space
where there is more underlying structure than we could ever hope to
completely document. What structure should we prioritize as the most
dominant or interesting in the data? %
} What is the single most important fact about a suspect? The second
most important fact? The third? And so on. Typically these decisions
are made on an ad hoc basis given the researcher's theoretical interests.
Here, the focus is in part learning the structure of the database
and so we need a principled definition of what makes an attribute
important, and a method for ranking attributes on that dimension.

I frame this as an unsupervised learning problem, where the task is
to learn rules and relationships between attributes that could be
used to distinguish a real observation from a synthetic randomly shuffled
version. The only way to tell a real observation from a randomly generated
one is to learn patterns of regularity and structure between attributes.
In this conception, an attribute is important if it conveys a great
deal of information about what other values a suspect's attributes
will take. The most important fact is the one that provides the most
information for inferring other facts. The least important fact is
the one that provides completely unique but orthogonal or potentially
random information.%
\footnote{Note that this is a reversal of the typical variable selection process,
where the goal is to better explain some outcome by removing redundant
information to produce a smaller number of uncorrelated explanatory
variables. In this multivariate setting, there is no single outcome
and the redundancies are the details of interest.%
}

The classifier I use for this task is an unsupervised random forest.%
\footnote{Implemented in the R package randomForestSRC \citep{ishwaran_random_2014}.%
} Random forests are an ensemble method that combines the predictions
of many individual base learners. The individual learners in this
case are fully grown binary decision trees, each fit to a different
random subset of attributes and random subset of observations \citep{breiman_random_2001}.%
\footnote{I employ 1000 trees, trying seven variables at each split, minimum
of one unique case at each split, and fully grown trees with no stopping
criteria. Splits with missing values are first determined using non-missing
in-bag observations, and then observations with that attribute missing
are randomly assigned to a child node.%
} In the supervised case, cut points for covariates are selected to
separate observations into increasingly homogeneous groups on some
outcome variable. In the unsupervised case, the random forest learns
to identify a genuine observation from a synthetic scrambled version
\citep{shi_unsupervised_2006}. This method works for both categorical
and continuous variables and is non-parametric, so there's no need
for prior knowledge of an underlying functional form.

The most useful information for this learning task is provided by
strong and regular relationships between variables, and so each decision
tree will tend to select variables with multiple strong interactions
earlier in the process, toward the root of the tree. Therefore, I
measure the importance of a variable as the average distance from
the root node to its maximal subtree (the earliest point in the tree
that splits on the variable) \citep{ishwaran_high-dimensional_2010}.
The interaction of two variables is captured by the depth of their
second-order maximal subtree (the distance from the root node of one
variable's maximal subtree within the maximal subtree of the other),
as trees tend to split on one variable and then soon split on a related
variable. I define a symmetric distance between two variables as the
sum of their second-order maximal subtree depths. This distance is
small when both variables tend to split close to the root, soon after
one another, and large if either splits late in the tree or far from
the other.%
\footnote{Summing the second-order maximal subtree depths is a stronger test
of interaction and is a novel innovation so far as the author is aware.%
} This approach provides two remarkable pieces of summary information
shown in Figure 2.

\begin{figure}[H]
\hfill{}\includegraphics[width=5in]{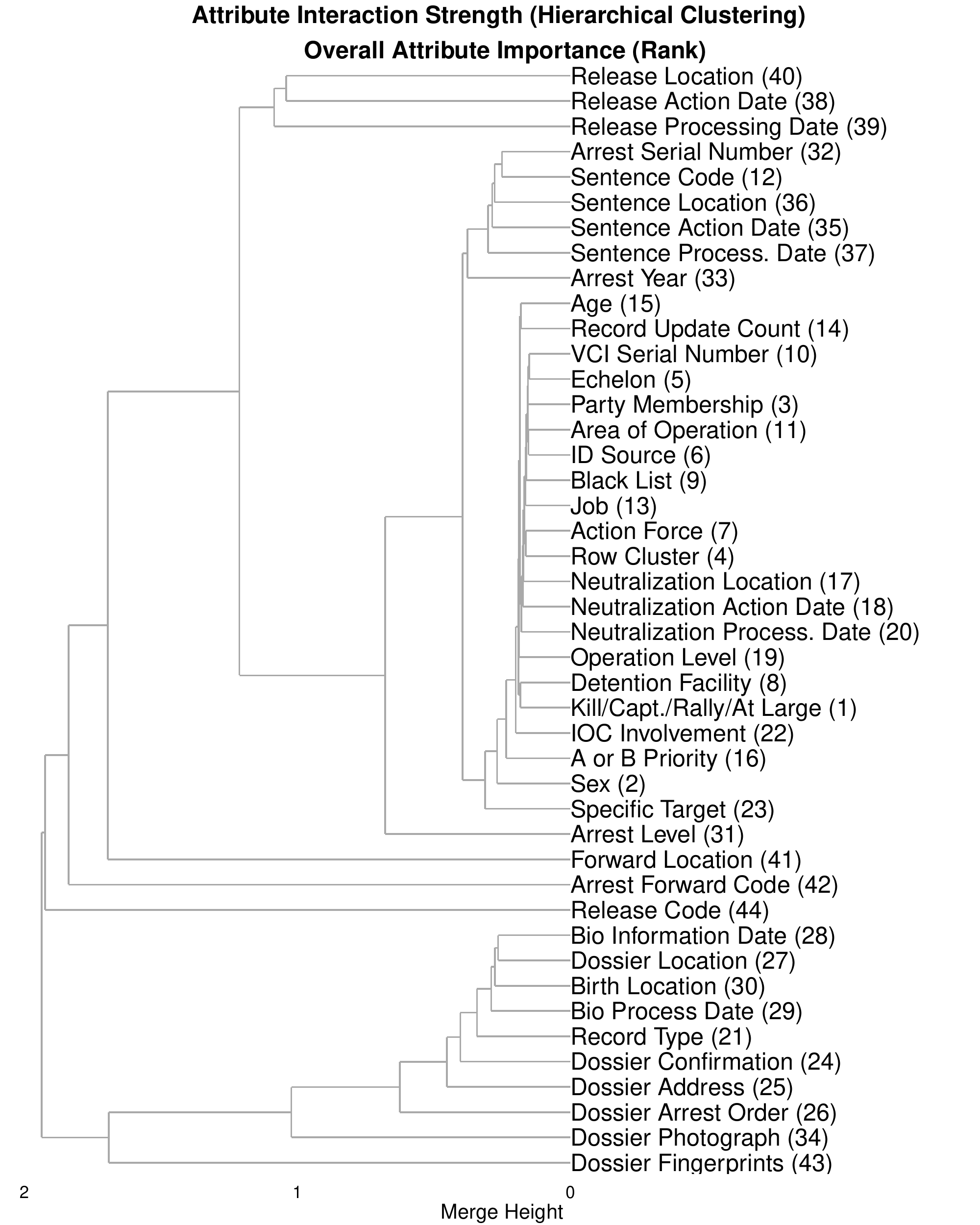}\hfill{}

\protect\caption{Clustering of attributes by strength of interaction with merges selected
to minimize Ward's distance (dendrogram). Rank order importance of
each attribute in terms of average maximal subtree depth in an unsupervised
learning task (unsupervsied random forest). Smaller rank means an
attribute was selected sooner in the random forest construction and
is thus more informative overall.}
\end{figure}

The first piece of summary information is a ranking of variables in
terms of how much information they convey about the entire dataset.
The answer of the question “Which fact should we start with?” is definitively
the fate of the suspect: still at large, killed, captured, or defector.
No other single attribute implies as much about the remaining details
as that one. The next most important is the suspect's gender, followed
by whether they were a party member, the kind of record as estimated
by the blockclustering above, the suspect's echelon, the source of
information used to ID the suspect, and so on.

The second is a pairwise distance between attributes in terms of how
much information their interaction conveys about the entire dataset.
Hierarchically clustering attributes on that distance reveals what
appears to be two mostly unconnected data generating processes: one
related to the creation of biographical records and dossiers, and
another related to the neutralization of suspects. The method has,
without prompting, correctly recovered the undocumented difference
between neutralization record attributes and biographical record attributes.
Justifying earlier concerns, the dossier attribute ``confirmed''
is flagged as being closely related to other administrative details
of dossiers and not the core demographic attributes of suspects or
the empirical process of targeting.%
\footnote{This is all the more amazing because the variable is incorrectly imputed
with values for the majority of rows in the dataset. The method has
correctly identified the subset of rows for which the variable takes
on meaningful values and has grouped it with related variables accordingly.%
}

The clustering also pinpoints the place to start the analysis: a core
group of 21 highly related and informative attributes relating to
the neutralization and demographics of suspects. They are flanked
by tangential groups of attributes relating to the sentencing of a
suspect, the release of a suspect, and the details of dossiers for
biographical records. There may be interesting structure within these
other groups of attributes, but they are mostly orthogonal to the
core outcomes of interest and so can be safely set aside for future
work.

Having selected a core group of attributes, the next question is whether
they can be further summarized by a simpler topology. The next two
sections unpack these attributes and tackle dimensionality reduction
with respect to two themes; the kinds of victims, and the kinds of
government operations. For that analysis, I weed the list further
to just 11 nominal demographic and neutralization attributes. I set
aside dates and locations. I exclude 5 attributes about the administrative
aspects of the dataset. And I single out two attributes, with a large
number of categories, for detailed analysis: the job of the suspect
and the government actor responsible for the neutralization.

\section{The Victims of Targeting}

Who were the Phoenix Program's victims? The program was charged with
dismantling nonmilitary rebel organizations in South Vietnam.%
\footnote{With captured documents and defector reports, GVN and U.S. intelligence
analysts mapped those organizations in great detail \citep[165]{conley_communist_1967}.%
} A full breakdown of suspect counts by organization, demographic attributes,
and final status is shown Table 3.

\begin{table}[H]
\hfill{}\begin{tabular}{llccccc}
\hline
& &  & \multicolumn{4}{c}{Outcome} \\ 
& &  & At Large & Killed & Captured & \multicolumn{1}{c}{Defector} \\ 
 &  & All & \% & \% & \% & \multicolumn{1}{c}{\%} \\ 
\hline
 & All  & $73,699$ & $\phantom{0}32$ & $\phantom{0}21$ & $\phantom{0}30$ & $\phantom{0}16$ \\
Organization & PRP  & $54,977$ & $\phantom{0}31$ & $\phantom{0}19$ & $\phantom{0}34$ & $\phantom{0}15$ \\
 & NLF  & $\phantom{0}7,147$ & $\phantom{0}33$ & $\phantom{0}13$ & $\phantom{0}28$ & $\phantom{0}26$ \\
 & Communist Orgs  & $11,315$ & $\phantom{0}39$ & $\phantom{0}33$ & $\phantom{0}13$ & $\phantom{0}15$ \\
Sex & Male  & $56,257$ & $\phantom{0}36$ & $\phantom{0}25$ & $\phantom{0}22$ & $\phantom{0}16$ \\
 & Female  & $16,910$ & $\phantom{0}20$ & $\phantom{00}6$ & $\phantom{0}57$ & $\phantom{0}17$ \\
Party & Full Member  & $22,443$ & $\phantom{0}48$ & $\phantom{0}29$ & $\phantom{0}13$ & $\phantom{0}10$ \\
 & Membership Unknown  & $29,470$ & $\phantom{0}37$ & $\phantom{0}19$ & $\phantom{0}31$ & $\phantom{0}12$ \\
 & Probationary Member  & $\phantom{0}7,042$ & $\phantom{0}17$ & $\phantom{0}26$ & $\phantom{0}37$ & $\phantom{0}20$ \\
 & Non-Member  & $14,743$ & $\phantom{00}6$ & $\phantom{0}10$ & $\phantom{0}51$ & $\phantom{0}33$ \\
Echelon & Hamlet  & $\phantom{0}9,858$ & $\phantom{0}25$ & $\phantom{0}24$ & $\phantom{0}24$ & $\phantom{0}26$ \\
 & Village  & $44,513$ & $\phantom{0}31$ & $\phantom{0}20$ & $\phantom{0}33$ & $\phantom{0}15$ \\
 & District  & $12,917$ & $\phantom{0}42$ & $\phantom{0}23$ & $\phantom{0}22$ & $\phantom{0}13$ \\
 & City/Prov/Reg/COSVN  & $\phantom{0}6,400$ & $\phantom{0}33$ & $\phantom{0}16$ & $\phantom{0}34$ & $\phantom{0}17$ \\
List & Most Wanted List  & $24,285$ & $\phantom{0}65$ & $\phantom{0}21$ & $\phantom{00}9$ & $\phantom{00}4$ \\
 & Target List  & $15,687$ & $\phantom{0}40$ & $\phantom{0}18$ & $\phantom{0}30$ & $\phantom{0}12$ \\
 & Most Active List  & $\phantom{0}8,136$ & $\phantom{0}23$ & $\phantom{0}15$ & $\phantom{0}51$ & $\phantom{0}11$ \\
 & Unknown  & $25,588$ & $\phantom{00}0$ & $\phantom{0}24$ & $\phantom{0}43$ & $\phantom{0}33$ \\
AorB & A  & $43,645$ & $\phantom{0}42$ & $\phantom{0}25$ & $\phantom{0}18$ & $\phantom{0}15$ \\
 & B  & $29,790$ & $\phantom{0}18$ & $\phantom{0}15$ & $\phantom{0}47$ & $\phantom{0}19$ \\
Age & [0,25]  & $15,500$ & $\phantom{0}19$ & $\phantom{0}14$ & $\phantom{0}47$ & $\phantom{0}21$ \\
 & (25,35]  & $14,858$ & $\phantom{0}37$ & $\phantom{0}22$ & $\phantom{0}25$ & $\phantom{0}16$ \\
 & (35,55]  & $27,225$ & $\phantom{0}35$ & $\phantom{0}14$ & $\phantom{0}31$ & $\phantom{0}20$ \\
 & (55,100]  & $\phantom{0}3,742$ & $\phantom{0}22$ & $\phantom{00}5$ & $\phantom{0}56$ & $\phantom{0}17$ \\
\hline 
\end{tabular}
\hfill{}

\protect\caption{Cross tabulation of demographic properties against outcome.}
\end{table}

Broadly, the political opposition to the Republic of Vietnam (GVN)
was organized into three groups. Political authority, command, and
resources flowed from North Vietnam into South Vietnam through the
communist political apparatus, the People's Revolutionary Party (PRP).
Indigenous popular support and participation was organized into the
subordinate National Liberation Front (NLF), also called the Viet
Cong. Together they constructed alternative administrative institutions
referred to as Communist Authority Organizations, such as the People's
Revolutionary Government (PRG), as well as a number of political organizations
designed to involve civilians outside of the communist party.

Together, and with overlapping and changing roles and capabilities,
these three organizations embodied foreign authority, popular participation,
and political institutions. Four-fifths of neutralizations were against
PRP positions with fewer directed toward more indigenous NLF and Communist
Organization positions. This is consistent with both the priorities
of the program and the timing in the war; the Post-Tet phase was more
externally driven by North Vietnam.

Members or supporters performing active roles of these organizations
were collectively known as the Viet Cong Infrastructure (VCI).%
\footnote{Military personnel serving in organizational roles, e.g. on Military
Affairs Committee, could qualify as VCI.%
} VCI were grouped into Class A VCI that were full or probationary
PRP members or leadership and command roles while class B VCI were
trained but voluntary members. Where appropriate, roles were replicated
at multiple levels of governance called echelons including the hamlet,
village, district, province, city, capital, region, and national level.

The broad pattern is one of a program that targeted large numbers
of low level suspects, a portfolio of targets that was bottom heavy.
Half of neutralizations were B level voluntary/support positions,
and a little more than half were previously unknown to security forces.
About a fifth of neutralizations were full party members, and about
a fifth of neutralizations were at the district or higher echelon.
It is unclear, however, whether this is disproportionate to the number
of actual rebels holding those positions. If the program selected
targets uniformly from the rebel population, these rates are probably
proportional to the share of those positions of all rebel members
during this period of the war.

There is a strong relationship between the demographics of suspects
and methods of targeting. In brief, more important suspects were fewer
in number but more directly targeted, either by having a file while
at large, or by being killed in an operation. Low level suspects were
less likely to have a file at large, and were much more likely to
be swept up as arrests or to walk in off the street as a defection.

The cross-tabulation in Table 3 shows this in terms of a single comparison
between demographics and the suspect's final status. Targeting was
gendered, with female suspects much less likely to be killed or to
be targeted as at large. Known party members were more likely to be
targeted at large, or killed, while suspects known to not be party
members were much more likely to simply be arrested or to defect.
The lower the suspect's echelon, the more likely they were simply
arrested or defected and the fewer targeted while at large.

The same is true for the level of prior suspicion against the suspect.
More prior suspicion is associated with more severe outcomes. Previously
unknown suspects tended to be arrested in the field or defectors,
not killed.%
\footnote{By definition, a previously unknown suspect (not on a black list)
did not have a biographical record (was not targeted at large).%
} Moving up the ladder of suspicion to the most active list, the target
list, and the most wanted list increases the chances that the suspect
is killed by an operation or targeted at large.%
\footnote{Note that the data speak to the probability of being under suspicion
given already being targeted. Estimating changes in the risk of being
targeted as a function of suspicion would require additional information
about the population of rebels overall.%
} The same pattern holds true for the A or B priority of the suspect's
position.

Visualizing the underlying pattern in just the bivariate case is already
somewhat overwhelming. Extending the analysis to the multivariate
case and developing a simplifying taxonomy is the task of dimensionality
reduction that I turn to next.

\subsection{A Taxonomy of Suspects}

Is there a simpler topology for understanding differences between
victims? I frame this as a dimensionality reduction problem where
nominal values for each of the categorical variables are mapped to
common latent dimensions. The method I use for this estimation is
Multiple Correspondence Analysis (MCA). MCA is a multivariate technique
analogous to Principal Components Analysis but for unordered categorical
data \citep{le_factominer:_2008}.

Let an $I\times Q$ matrix represent the values for each individual
$i$ and attribute $q$ with $K_{q}$ possible values for each attribute
and $K$ total possible values. This matrix is then converted to an
$I\times K$ disjunctive table (dummy variables for each level of
each variable). Rarely used categories disproportionately influence
the construction of these dimensions, so I suppress both rare and
missing values.%
\footnote{This is another motivation for carefully studying missingness in the
database. The main source of variation in the database is technical,
the difference between different kinds of records. I manually suppress
missing values and purely administrative variables so that the estimated
components reflect only the substantive empirical variation between
attributes.%
} 

I use a variant of the algorithm called Specific Multiple Correspondence
Analysis that correctly calculates partial distance between points
given levels were dropped \citep{roux_multiple_2009}.%
\footnote{Implemented in the R package soc.ca.%
} It decomposes the disjunctive table into principle axis representing
latent dimensions, points for individuals in that reduced space, and
points for each attribute value in the same space. The result is a
geometric interpretation of the originally categorical data where
suspects and attributes are all now projected into a smaller number
of continuous dimensions.

The core variation of the dataset is well summarized by a few latent
dimensions, with the first two principal axis accounting for 74\%
of total variation (inertia).%
\footnote{In total 18 dimensions account for 100\% of variation.%
} They are summarized in Table 4. The first dimension (56\%) reflects
a clear demographic concept of the suspect's importance to the targeting
program. At one extreme are unimportant, previously unknown, low level
volunteers, often caught in large raids. At the extreme are high level,
full party members, that are on the most wanted list, but usually
remain at large.

The second dimension (18\%) reflects the method of neutralization
used. At one extreme are killings, sometimes targeting specific individuals,
in operations directed by the local intelligence office. At the other
extreme are defections, that required no previous effort by an intelligence
office, where the identity was confirmed by the suspects own confession.
In between lie arrests which share aspects of both kinds of targeting.

\begin{table}[H]
\hfill{}%
\begin{tabular}{|c|c>{\centering}p{0.25in}c|ccc|}
\cline{2-7} 
\multicolumn{1}{c|}{} & Dimension 1 (56\%) & Ctr. & Coord. & Dimension 2 (18\%) & Ctr. & Coord.\tabularnewline
\hline 
\multirow{6}{*}{(+)} & At Large & 13.7 & 1.18 & Killed & 14.4 & 1.21\tabularnewline
 & Most Wanted List & 10.5 & 1.03 & Order of Battle for ID & 13.3 & 1.45\tabularnewline
 & Full Party Member & 5.6 & 0.78 & Specific Target & 8.8 & 1.05\tabularnewline
 & A Priority & 4.0 & 0.47 & Result of IOCC Information & 7.7 & 0.71\tabularnewline
 &  &  &  & Subsector/District Level Op. & 6.9 & 0.57\tabularnewline
 &  &  &  & Directed by IOCC & 5.4 & 0.98\tabularnewline
\hline 
 & Non-specific Target & 8.3 & -0.79 & Defector & 10.9 & -1.19\tabularnewline
\multirow{5}{*}{(-)} & Captured & 8.2 & -0.95 & Confession for ID & 7.1 & -0.98\tabularnewline
 & Unknown (No List) & 6.2 & -0.77 & No IOCC Involvement & 3.0 & -0.7\tabularnewline
 & B Priority & 5.8 & -0.69 &  &  & \tabularnewline
 & Subsector/District Level Op. & 5.4 & -0.63 &  &  & \tabularnewline
 & Result of IOCC Information & 4.3 & -0.66 &  &  & \tabularnewline
 & Non-Party Member & 4.0 & -0.81 &  &  & \tabularnewline
 & Agent/Informer for ID & 4.0 & -0.78 &  &  & \tabularnewline
 & Female & 3.6 & -0.73 &  &  & \tabularnewline
 & Age {[}0,25{]} & 3.0 & -0.69 &  &  & \tabularnewline
\hline 
\end{tabular}\hfill{}

\protect\caption{The first two dimensions of demographic and operations attributes
estimated with multiple correspondence analysis for all suspects.
Contribution and coordinates of specific values shown for attributes
with above average contribution to each dimension.}
\end{table}

This provides a clean language for describing civilian targeting in
terms of just two concepts. First, there are suspects the government
wishes it could target and those that it actually targets in practice.
Second, of those it targets, there is a spectrum ranging from suspects
that tend to defect, suspects that tend to be arrested or captured
out in the field, and suspects that tend to be assassinated (or who
die in battle but the program claims credit). A map of each value
in the two dimensions is shown in Figure 3.

\begin{figure}[H]
\hfill{}\includegraphics[width=5in]{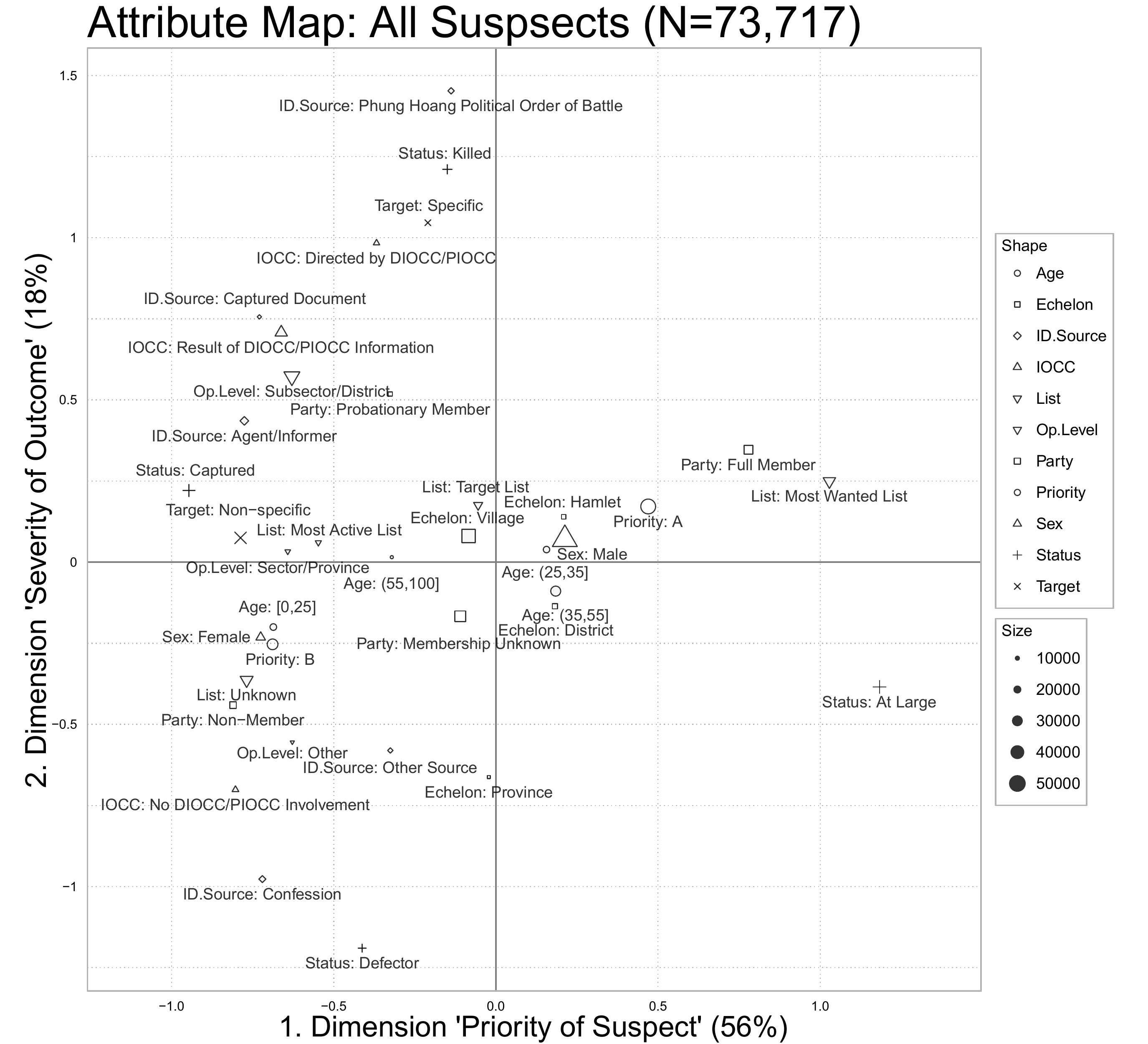}\hfill{}

\protect\caption{Attribute values for all suspects projected into two dimensions with
multiple correspondence analysis.}
\end{figure}

For studies that use counts of rebel or civilian deaths as a dependent
variable, this shows that those raw counts could be driven by changes
in at least three different underlying dynamics: (1) the intensity
of the war, growing or shrinking the size of the target list or the
number of operations looking for suspects not on a list; (2) changes
in effectiveness, neutralizing more or fewer known targets already
on the list; or (3) changes in tactics, using more killings than arrests
or more defections preempting killings, etc. Shifts along any of these
dimensions could produce changes in total body counts or the portfolio
of observed violence (e.g. ratios of civilian to military deaths).
There is currently little in the way of theoretical expectations for
how interventions should interact with each of these dimensions, much
less how those interactions should aggregate into changes in total
observed levels of violence.

\subsection{The Jobs Held by Suspects}

As an external check of validity, I compare the estimated position
of each suspect along the dimensions of targeting to a description
of the job they held. If the dimensions are correct, and useful, then
jobs with similar functions ought to be more similar to each other
in terms of targeting. I find that suspects with similar jobs, as
described by third party sources, do in fact have similar demographic
attributes and similar targeting behavior by the government. If the
underlying data were faked or entered with error, they were at least
doing it in a consistent and creative way.

Each suspect is tagged with one of 485 specific jobs, coded according
to a standardized official schema called the Greenbook. Each jobs
is nested within increasingly large departments called elements, subsections,
sections, and the three main branches. I focus on the section level
of aggregation. The location of each section along the dimensions
of targeting is estimated by including the section attribute as a
noncontributing covariate in the Multiple Correspondence Analysis
introduced earlier. A map of sections along the two dimensions of
targeting is shown Figure 4.

\begin{figure}[H]
\hfill{}\includegraphics[width=5in]{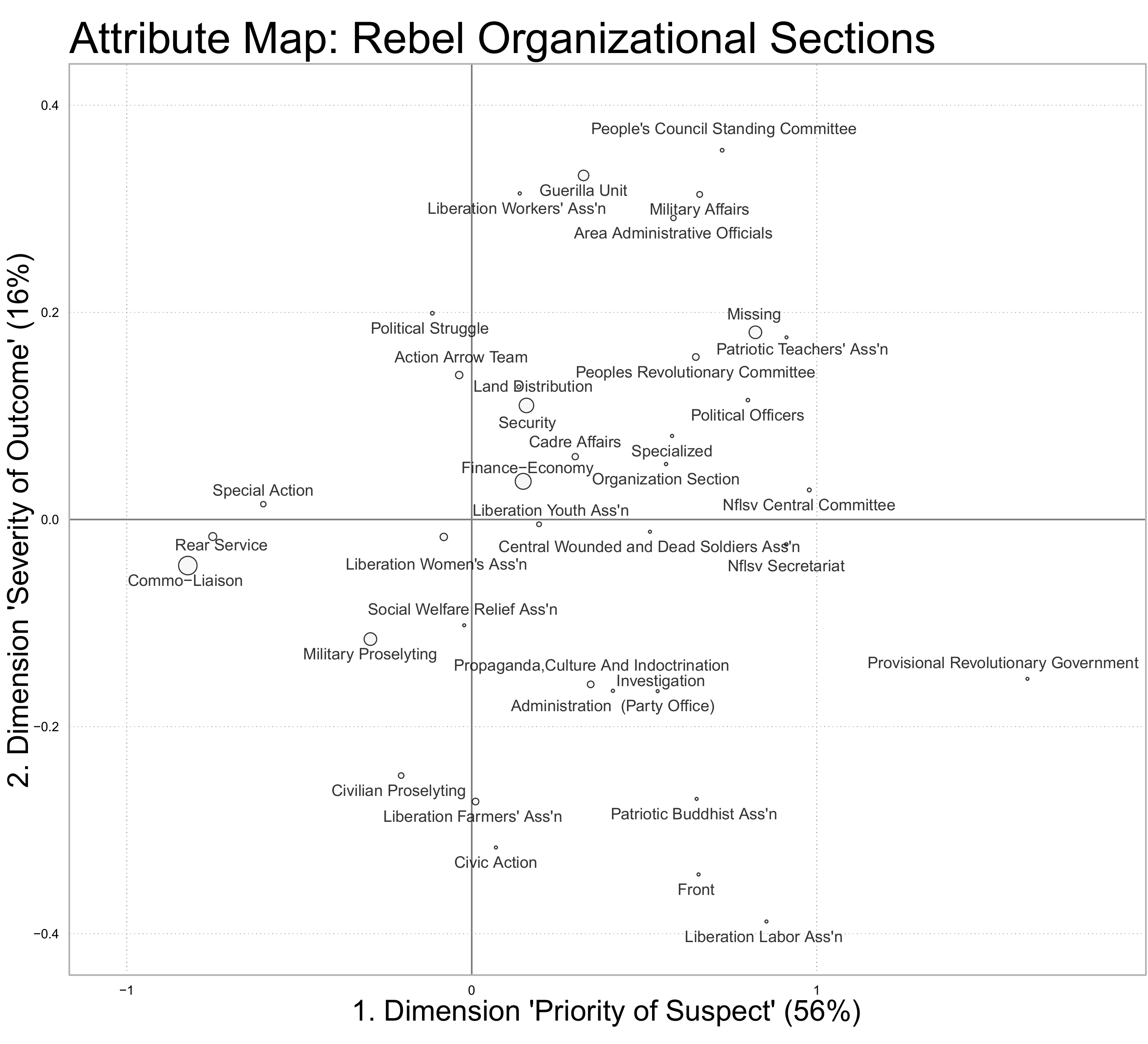}\hfill{}

\protect\caption{Map of rebel sections projected into the two dimensions of suspect
importance and severity of outcome.}
\end{figure}

Most of the variation across sections is on the first dimension of
priority (horizontal axis). At one extreme on the far left are low
level logistics related sections like the Commo-Liaison and Rear Service
sections, where suspects were rarely targeted at large and mostly
swept up as arrests. At the other extreme, on the right, are high
level leadership positions like the NLF Central Committee and the
Provisional Revolutionary Government where almost every suspect was
most wanted but still at large. Along the second dimension of severity
of outcome (vertical axis), some sections were likely to be specifically
targeted or killed, e.g. Guerilla Units, Military Affairs Section,
or Area Administrative Officials. At the other extreme some groups
were much more likely to defect, e.g. the Medical Section, the Frontline
Supply Council, or the Western Highlands Autonomous People's Movement.

Next I cluster the sections according to their distance along the
targeting dimensions.%
\footnote{I calculate euclidean distance on the first three dimensions which
account for over 80\% of the variation.%
} The 30 sections with 100 or more suspects are described in Table
5. They are arranged hierarchically using Ward's method by their proximity
in biographical dimensions estimated with MCA above \citep{ward_hierarchical_1963}.
Each section is provided with a brief description based on their functions
as outlined by U.S. intelligence \citep{combined_intelligence_center_vietnam_vci_1969}.

\begin{table}[H]
\hfill{}%
\begin{tabular}{>{\raggedright}p{0.9in}>{\raggedright}p{1.5in}>{\raggedleft}p{0.3in}>{\raggedright}p{1.25in}>{\raggedright}p{0.25in}}
\cline{2-5} 
 & Section & N & Description & Org.\tabularnewline
\cline{2-5} 
\multirow{23}{0.9in}{{\small{}\includegraphics[width=1in,height=5.01in]{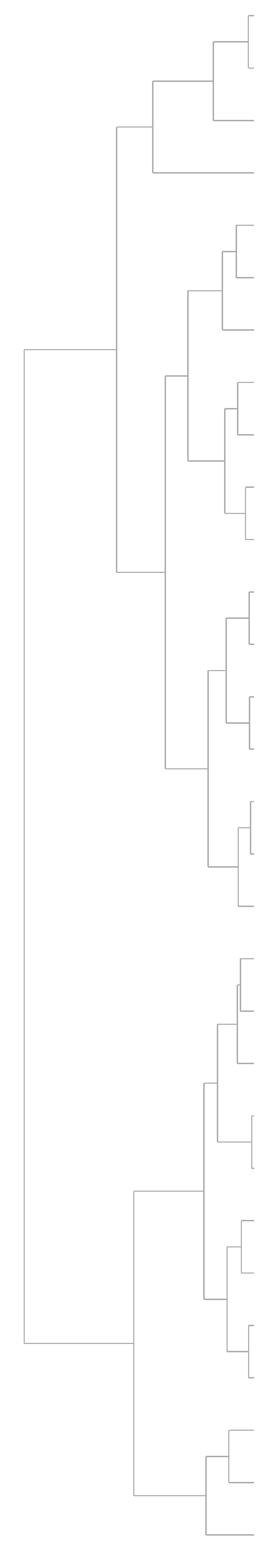}}} & Liberation Farmers' Ass'n  & 2,230 & Mass Org. Local & NLF\tabularnewline
 & Cadre Affairs  & 2,007 & Intel, Proselytizing & PRP\tabularnewline
 & Guerilla Unit & 4,593 & Armed local forces & Org\tabularnewline
 & Military Affairs & 1,671 & Coordinate Guerrillas & PRP\tabularnewline
 & People's Council & 447 & Administration & Org\tabularnewline
 & Area Adm. Officials & 1,406 & Administration & Org\tabularnewline
 & Peoples Rev. Comm. & 2,202 & Administration & PRP\tabularnewline
 & Organization Section & 188 & Administration & PRP\tabularnewline
 & Specialized & 126 &  & NLF\tabularnewline
 & Nflsv Secretariat & 216 & NLF Leadership & NLF\tabularnewline
 & Nflsv Central Committee & 594 & NLF Leadership & NLF\tabularnewline
 & Political Officers & 337 & PRP Leadership & PRP\tabularnewline
 & Liberation Workers' Ass'n  & 137 & Mass Org, Urban & NLF\tabularnewline
 & Land Distribution  & 663 &  & PRP\tabularnewline
 & Security & 6,990 & Intel., Police, Justice & PRP\tabularnewline
 & Party Office & 163 & Administration  & PRP\tabularnewline
 & Culture-Indoctrination & 2,210 & Propaganda & PRP\tabularnewline
 & Finance-Economy  & 7,845 & Logistics, Taxes, Food & PRP\tabularnewline
 & Liberation Youth Ass'n  & 1,050 & Mass Org, Youth & NLF\tabularnewline
 & Action Arrow Team & 2,665 & Mobile Security & Org\tabularnewline
 & Political Struggle  & 443 &  & \tabularnewline
 & Liberation Women's Ass'n  & 2,654 & Mass Org. Women. & NLF\tabularnewline
 & Military Proselyting & 5,848 & Turn GVN soldiers & PRP\tabularnewline
 & Medical & 3,387 & Public/Civil health & PRP\tabularnewline
 & Civilian Proselyting  & 1,538 & Party Recruiting & PRP\tabularnewline
 & Frontline Supply Council & 681 & Logistics. & PRP\tabularnewline
 & Production & 1,375 & Rear Production & PRP\tabularnewline
 & Special Action & 1,425 & Sappers & PRP\tabularnewline
 & Commo-Liaison  & 9,425 & Logistics, Routes & PRP\tabularnewline
 & Rear Service  & 3,037 & Logistics, Military & PRP\tabularnewline
\cline{2-5} 
\end{tabular}\hfill{}

\protect\caption{Organizational sections with over 100 suspects. Hierarchical clustering
with Ward's distance shown in dendrogram on the left.}
\end{table}

The clustering has recovered groups of sections with similar functions,
arranged into roughly four themes. The fighting themed cluster contains
four large sections including the Guerrilla Unit, Military Affairs,
Cadre Affairs, and the Liberation Farmers Association. All operated
at the low village or hamlet level and were high risk in terms of
chance of being killed.

A leadership themed cluster includes seven small groups with positions
that operated at the village, district, or higher echelon, and were
more likely to be targeted while at large or killed if neutralized.
This includes the NLF Secretariat, the NLF Central Committee, Area
Administration Officials, etc.

There are two village administration themed clusters. The first leans
toward higher priority and more violent outcomes. It includes, Political
Officers, the Security Section, the Party Office, the Culture and
Indoctrination, and the Finance-Economy Section. This captures the
main justice, propaganda, and tax collection infrastructure at the
district and village level. The second administration themed cluster
includes village level Womens' and Youth associations, the Military
and Civilian Proselyting Sections, and sections responsible for medical
and food production. This captures very local leadership and organization
at the village level.

A final logistics themed cluster includes three sections, Special
Action, Commo-Liaison, and Rear Service. Suspects in these sections
were much more likely to be captured, of very low priority in terms
of not being party members, not on a wanted list, B level positions,
and women.

This set of clusters provides a way to think about targeting the different
components of an insurgency: a fighting component, a leadership component,
a day to day administration component, and a logistical tail. Both
the demographics of suspects and the targeting methods of the government
vary systematically across these different components. This is important
for studies that use violence over time as a dependent variable. If
any of these groups change in size or in level of activity, it will
change the aggregate body count in potentially unpredictable ways.

\section{The Methods of Targeting}

Killings and arrests required the government to launch an operation,
and defections required a receiving government actor or office. Which
government actors conducted those operations and what methods did
they use? When a suspect is neutralized, the details of the circumstance
or operation leading to their neutralization were recorded. The details
of each neutralization are cross-tabulated against outcomes in Table
6.

\begin{table}[H]
\hfill{}\begin{tabular}{llcccc}
\hline
& &  & \multicolumn{3}{c}{Outcome} \\ 
& &  & Killed & Captured & \multicolumn{1}{c}{Defector} \\ 
 &  & All & \% & \% & \multicolumn{1}{c}{\%} \\ 
\hline
 & All  & $49,774$ & $\phantom{0}31$ & $\phantom{0}45$ & $\phantom{0}24$ \\
Source & Agent/Informer  & $16,296$ & $\phantom{0}37$ & $\phantom{0}56$ & $\phantom{00}7$ \\
 & Captured Document  & $\phantom{0}4,428$ & $\phantom{0}44$ & $\phantom{0}45$ & $\phantom{0}11$ \\
 & Confession  & $11,633$ & $\phantom{00}7$ & $\phantom{0}42$ & $\phantom{0}52$ \\
 & Order of Battle  & $\phantom{0}9,942$ & $\phantom{0}50$ & $\phantom{0}41$ & $\phantom{00}9$ \\
 & Other Source  & $\phantom{0}7,249$ & $\phantom{0}22$ & $\phantom{0}31$ & $\phantom{0}47$ \\
Level & DTA/Other/Region  & $\phantom{0}3,835$ & $\phantom{0}30$ & $\phantom{0}43$ & $\phantom{0}27$ \\
 & Sector/Province  & $\phantom{0}6,880$ & $\phantom{0}27$ & $\phantom{0}58$ & $\phantom{0}15$ \\
 & Subsector/District  & $33,296$ & $\phantom{0}37$ & $\phantom{0}50$ & $\phantom{0}13$ \\
IOCC & No Involvement  & $\phantom{0}9,637$ & $\phantom{0}29$ & $\phantom{0}46$ & $\phantom{0}25$ \\
 & Result of Info  & $24,127$ & $\phantom{0}38$ & $\phantom{0}52$ & $\phantom{00}9$ \\
 & Directed by  & $\phantom{0}8,785$ & $\phantom{0}38$ & $\phantom{0}59$ & $\phantom{00}3$ \\
Specific & Non-specific  & $32,467$ & $\phantom{0}31$ & $\phantom{0}49$ & $\phantom{0}20$ \\
 & Specific  & $12,630$ & $\phantom{0}43$ & $\phantom{0}49$ & $\phantom{00}8$ \\
\hline 
\end{tabular}
\hfill{}

\protect\caption{Properties of operations across neutralization outcomes.}
\end{table}

The tabulations show a program with a large base of incidental arrests
and killings in the course of regular operations topped with a sizable
number of direct planned strikes against specific targets. A full
third of killings and captures were suspects target by an operation,
often an ambush along a route or a raid. About a third were from operations
directed by an IOCC. As noted before, more than half of killed or
captured suspects were already previously listed on a blacklist.

The identity of a suspect had to be confirmed at the time of neutralization.
For previously unidentified suspects, the source of ID at the time
of neutralization may have been the source that led them to be a suspect
in the first place. For suspects already on a blacklist or listed
in the Political Order of Battle, there was some unobserved process
by which evidence was collected, leading to the initial suspicion.
The majority were either identified by another civilian (an agent
or informer) or they were said to have confessed. Others were confirmed
by material evidence like documents captured on their person or identifying
them by name. Some were confirmed against descriptions in the Political
Order of Battle (OB). About 12\% were identified as ``other'' explained
in a written comment on the back of the worksheet and not recorded
here.

\subsection{A Taxonomy of Targeting Operations}

Neutralizations can also be well summarized by a simpler typology.
I fit the same specific multiple correspondence model as before to
just the subset of observations resulting in arrest or killing. The
first principal axis accounts for 59\% of the variation and, as before,
reflects the level of priority of the suspect. At one extreme are
low level incidental captures and at the other killings against priority
targets. To a lesser degree it also captures the level of premeditation,
as high priority targets were more likely to be the specific target
of an operation and the target of an operation planned and directed
by an IOCC.

\begin{table}[H]
\hfill{}%
\begin{tabular}{|c|c>{\centering}p{0.25in}c|ccc|}
\cline{2-7} 
\multicolumn{1}{c|}{} & Dimension 1 (59\%) & Ctr. & Coord. & Dimension 2 (15\%) & Ctr. & Coord.\tabularnewline
\hline 
\multirow{7}{*}{(+)} & Most Wanted List & 13.3 & 1.37 & No IOCC Involvement & 19.0 & 1.34\tabularnewline
 & Full Party Member & 10.9 & 1.11 & Other Source for ID & 14.1 & 1.59\tabularnewline
 & A Priority & 7.8 & 0.66 & Sector/Province Level Op. & 8.0 & 0.96\tabularnewline
 & Killed & 7.8 & 0.73 & Other Level Op. & 7.7 & 1.57\tabularnewline
 & Order of Battle for ID & 6.9 & 0.9 & Province Echelon & 3.7 & 1.19\tabularnewline
 & Specific Target & 5.5 & 0.71 & Unknown (No List) & 3.2 & 0.36\tabularnewline
 &  &  &  &  &  & \tabularnewline
\hline 
 & B Priority & 7.9 & -0.67 & Result of IOCC Information & 9.6 & -0.55\tabularnewline
\multirow{4}{*}{(-)} & Female & 5.8 & -0.76 & Captured Document for ID & 5.3 & -0.95\tabularnewline
 & Captured & 5.4 & -0.51 & Subsector/District Level Op. & 4.7 & -0.33\tabularnewline
 & Age {[}0,25{]} & 3.3 & -0.61 & Most Active List & 3.0 & -0.62\tabularnewline
 & Confession for ID & 3.0 & -0.76 & Target List & 2.9 & -0.51\tabularnewline
\hline 
\end{tabular}\hfill{}

\protect\caption{The first two dimensions representing demographic and operations related
attributes estimated with multiple correspondence analysis for only
observations resulting in killing or capture. Contribution and coordinates
of specific values shown for attributes with above average contribution
to each dimension.}
\end{table}

The second principal axis accounts for 15\% of the variation and reflects
the domain of the operation. On one end are operations that found
VCI and reported them retroactively to the intelligence infrastructure
for documentation. These operations typically had no IOCC involvement,
were carried out by more conventional forces, and at the province
or sector level. At the other extreme are operations carried out by
Phoenix related forces against targets known about beforehand. These
operations were at the subsector or district level, against village
echelon level targets, who were on the most active or target black
lists, and benefited from information provided by the IOCC. The map
of each attribute value along these two dimensions is shown in Figure
5.

\begin{figure}[H]
\hfill{}\includegraphics[width=5in]{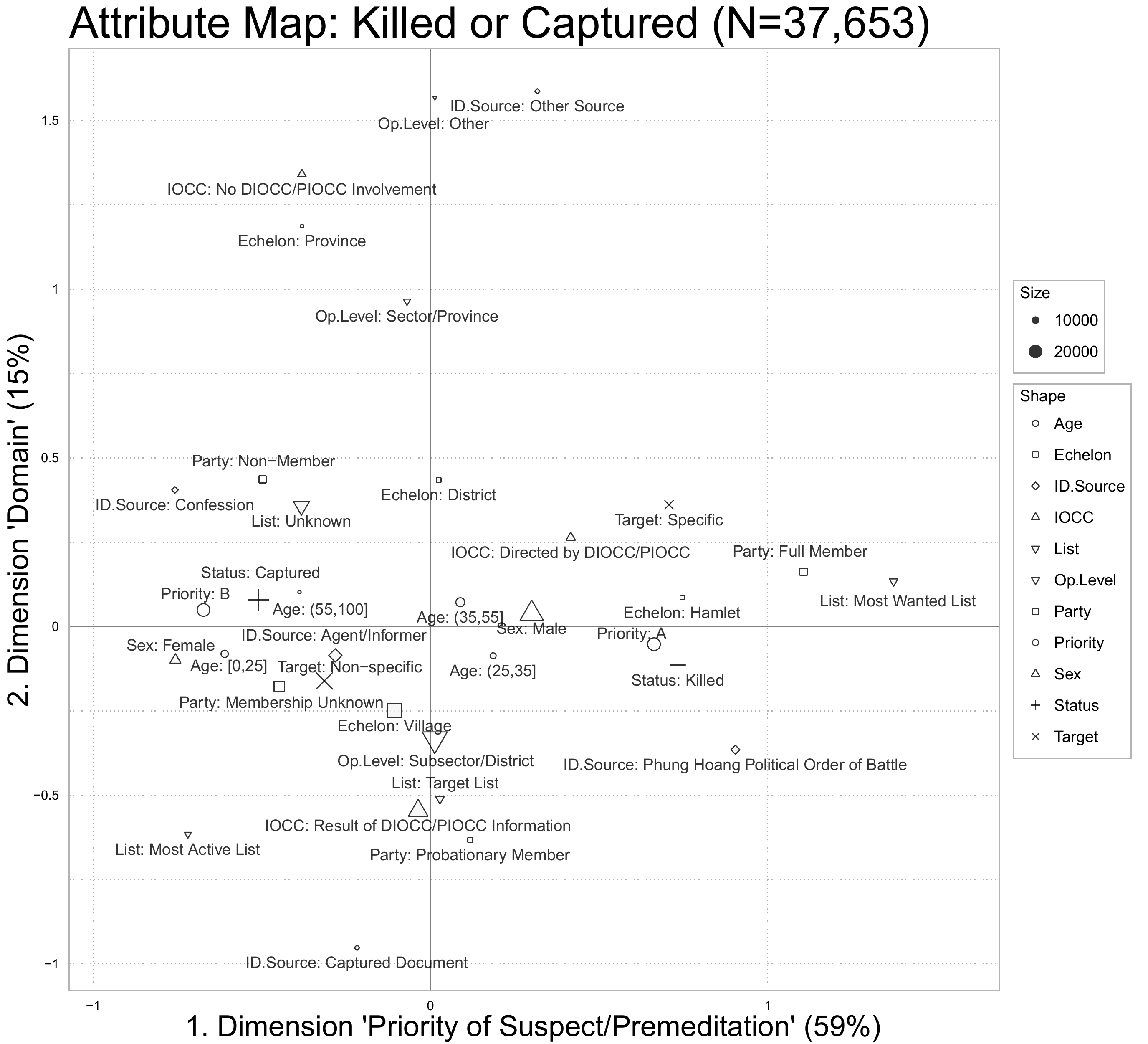}\hfill{}

\protect\caption{Attribute values for just killing and captures projected into two
dimensions with multiple correspondence analysis.}
\end{figure}

\subsection{The Perpetrators of Targeting}

There were 16 different organizations reported as the government actor
in the neutralization of suspects. As an additional external check
of validity, Figure 6 shows the position of the actors along the two
dimensions of priority and domain. If the underlying data are accurate
and well summarized by these dimensions, then actors with similar
functions ought to be similar to each other in terms of victims and
tactics.

\begin{figure}[H]
\hfill{}\includegraphics[width=5in]{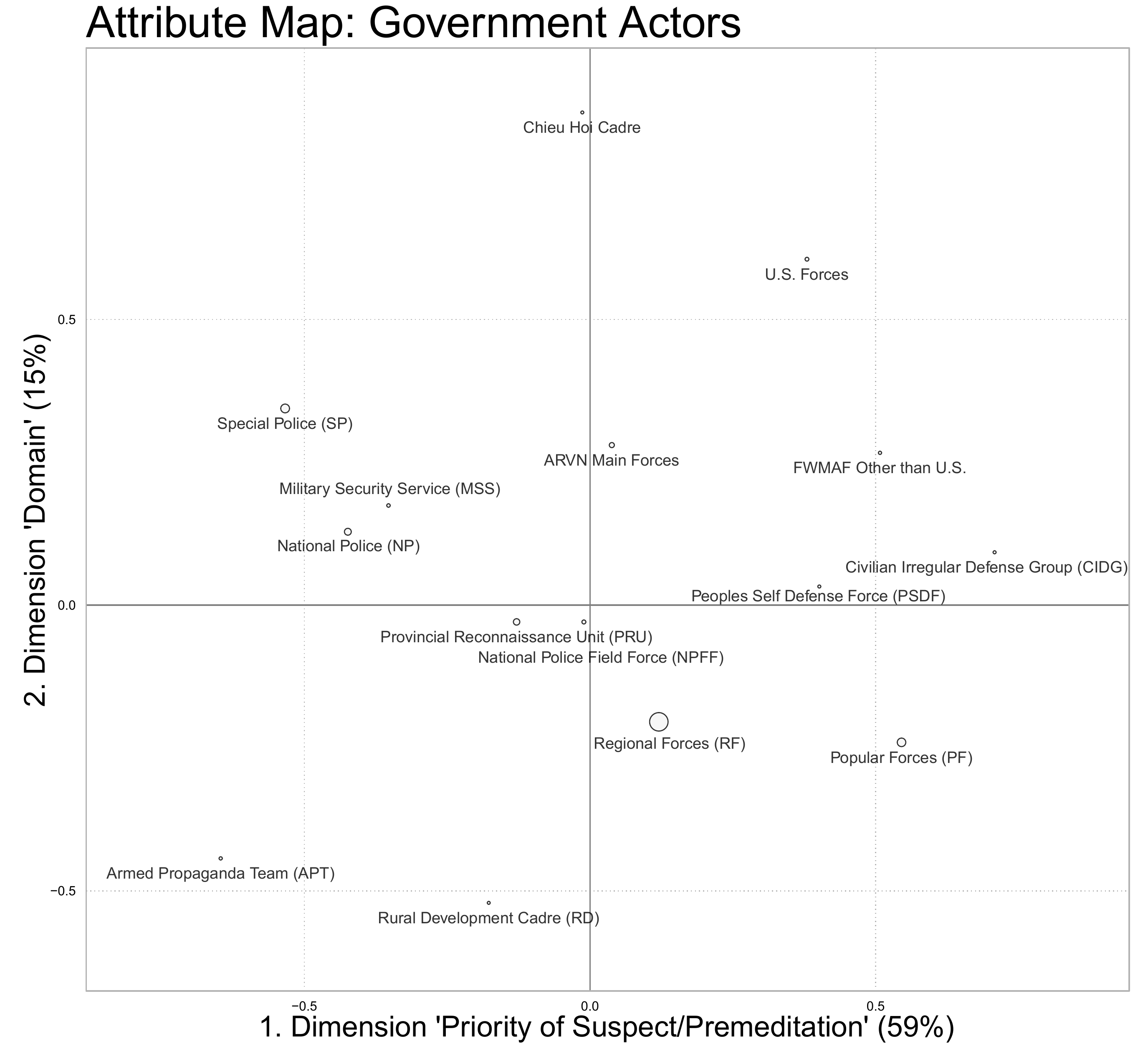}\hfill{}

\protect\caption{Government actors responsible for killings and captures projected
into the two dimensions of importance and domain.}
\end{figure}

A clustering of actors along the dimensions of operations are shown
in Table 8. On paper, the official Phoenix Forces were the National
Police, the Provincial Reconnaissance Units, Rural Development Cadre,
Civilian Irregular Defense Group, and the Armed Propaganda Team (APT).
In practice, the tent poles of the Phoenix Program were the Popular
Forces, Regional Forces and Special Police who made up two-thirds
of all killings and captures.

\begin{table}[H]
\hfill{}%
\begin{tabular}{>{\raggedright}p{0.4in}l>{\raggedleft}p{0.4in}l}
\cline{2-4} 
 & Actor & N & Description\tabularnewline
\cline{2-4} 
\multirow{14}{0.4in}{{\small{}\includegraphics[width=0.5in,height=2.3in]{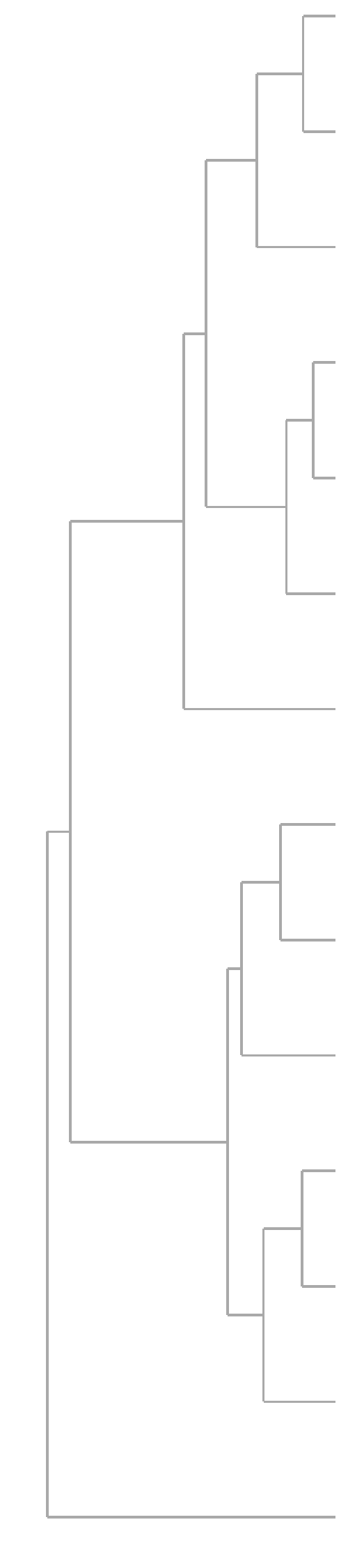}}} & Special Police (SP) & 5,375 & Urban Police\tabularnewline
 & National Police (NP) & 3,604 & Urban/Suburban Police\tabularnewline
 & Military Security Service (MSS) & 561 & Counter-intelligence\tabularnewline
 & Provincial Reconnaissance Unit (PRU) & 3,190 & Mobile Special Forces\tabularnewline
 & National Police Field Force (NPFF) & 1,065 & Mobile Rural Policing\tabularnewline
 & Army of the Republic of Viet Nam (ARVN) & 1,919 & Regular GVN Military\tabularnewline
 & Armed Propaganda Team (APT) & 310 & Mobile Cultural Team\tabularnewline
 & Regional Forces (RF) & 14,356 & District/Province Paramilitary\tabularnewline
 & Popular Forces (PF) & 5,195 & Hamlet/Village Paramilitary\tabularnewline
 & Civilian Irregular Defense Group (CIDG) & 146 & Irregular Militia\tabularnewline
 & Peoples Self Defense Force (PSDF) & 111 & Irregular Militia\tabularnewline
 & Free World Military Assistance Force & 308 & Allied, South Korea\tabularnewline
 & U.S. Forces & 899 & Regular U.S. Military\tabularnewline
 & Other & 539 & \tabularnewline
\cline{2-4} 
\end{tabular}\hfill{}

\protect\caption{Government actors responsible for neutralizations. Grouped according
to similarity on operation properties estimated with multiple correspondence
analysis. Dendrogram shows hierarchical clustering using Ward's method.}
\end{table}

It reveals four small clusters. The first cluster includes urban and
suburban police organizations that were much more likely to arrest
than kill. This is likely because they were both in areas of greater
government control where contestation was less violent overall and
because they were more likely to document their arrests than less
institutionalized forces.

The second cluster includes mobile forces who operated in areas of
weaker government control but still had close ties to the Phoenix
Program in terms of intelligence sharing and reporting. Provincial
Reconnaissance Units (PRU), for example, were specially designed units
for the Phoenix Program who had the highest rate of neutralizations
per fighter \citep[210]{thayer_war_1985}.

The third cluster contains paramilitary forces, Regional Forces were
responsible for routes and intersections while Popular Forces were
responsible for village and hamlet defense. Their neutralizations
were the most violent and numerous. This is because of where they
operated (in more contested areas), their high numbers (having more
manpower in more places than police forces), and their tactics (equipped
and trained for defense and attack rather than regular policing).

A final cluster represents conventional forces: the U.S. and Free
World Military Assistance forces (primarily South Korean). These forces
reported few suspects to the national database, likely because of
parallel reporting mechanisms and also because they were engaged in
larger more conventional fighting.

These clusters provide a way to think about targeting as a function
of different kinds of government forces: regular police, expeditionary
forces, paramilitary forces, and conventional forces. Each employs
different tactics, in a different environment, with a different portfolio
of victims, and different incentives and capabilities to report back
statistics. This is particularly relevant to the study of aggregate
levels of violence. Over the course of a conflict, the size and level
of activity of these four groups will change. Forces are raised, units
move around the country, and police and paramilitary forces are extended
to newly secured communities. Each of these will impact the amount
of violence committed and the number of casualties reported in a given
area over a given period of time.

\section{Conclusion}

The Phoenix Program provides an unusually clear view of a large wartime
government targeting effort. In the aggregate, it provides an example
of a typically mixed targeting program. Most processed neutralizations
were of low priority targets, while occasionally the program had the
intelligence, or good luck, to launch operations against high-profile
targets. This pattern is a result of the fundamental feature of civil
war, an inability to easily separate rebels from neutral civilians.

That said, based on the portion of targeting recorded in the national
database, the government did track and target a large number of suspects
with verifiable links. In the same way that policing is primarily
about deterring illegal behavior through small risks of punishment,
the Phoenix Program offered a credible risk to rebels who might have
otherwise operated openly or civilians who were on the fence about
joining.

There is a strong connection between a suspect's demographics, their
position in rebel organizations, the kind of government actor that
would target them, the methods that would be used, and their ultimate
fate. A combination of dimensionality reduction and clustering suggests
a few simplifying ways to describe that connection. Suspects vary
in priority to the government in terms of who it wishes it could target
and who it targets in practice. The outcomes for suspects vary in
severity ranging from voluntary defection to death in the field. Operations
vary in the priority of the suspect, arresting large numbers of unimportant
suspects in sweeps and launching premeditated operations to kill more
important suspects. Operations also vary across domains. Some operations
are carried out far away from intelligence and police infrastructure
and are underreported in official statistics. Other operations are
carried out in clear view, often using previous intelligence and regularly
reporting back for inclusion in official statistics.

The data also reveal clear organizational differences between different
government actors and different rebel sections. Rebel sections have
functions that fall into groups related to fighting, high level leadership,
low level administration, and logistical support. Government actors
fall into groups of regular police, expeditionary forces, paramilitary
forces, and conventional forces. Each kind of organizational subdivision
has a distinct signature in terms of types of civilians involved and
the types of targeting methods employed.

One motivation for moving toward bigger (wider) data in conflict studies
is that they reveal these underlying dimensions, organizational types,
and processes which typically get relegated to an error term. These
results should be a cautionary tale for analysis based on raw event
counts, often drawn from newspapers or similarly shallow reporting.
Targeting in the Vietnam War was very high dimensional. An analyst
could reach dramatically different conclusions about outcomes by truncating
the sample to just killings, by omitting information about the suspect's
position or the government actor committing the violence, or by missing
important details about how the institution created records and aggregated
them into a final dataset.

With data this detailed, the analysis provided here is just the tip
of the iceberg. I have shown ways to identify and explore the main
sources of variation in a database, but there are many other places
in the data to look for interesting structure. There are three that
come immediately to mind. The first is exploring the spatial and temporal
structure of data. The Vietnam War varied from province to province
and often from village to village, which should have clear implications
for how civilians were treated. The second is how neutralizations
were related to one another. I have treated neutralizations as isolated
events, but in reality they were often part of larger operations.
There is enough detail on locations and timing to aggregate individual
observations into a larger event level analysis. Finally, the structure
of rebel organizations is an entire field of study on its own, and
the detailed data on rebel jobs, locations, and demographics can provide
a remarkable map of Viet Cong and North Vietnamese organization across
South Vietnam.

\bibliographystyle{authordate1}
\bibliography{BookChapter}

\end{document}